\documentclass[pre,showpacs,showkeys,preprint]{revtex4}
\usepackage{bbm}
\usepackage{enumerate}
\usepackage{graphicx}
\usepackage{amssymb,amsmath,amsthm}

\newtheorem{theorem}{Theorem}
\newtheorem{lemma}{Lemma}
\newtheorem{cor}{Corollary}
\theoremstyle{definition}
\newtheorem{defn}{Definition}

\begin{document}

\title{Scale-Invariant Cellular Automata and Self-Similar Petri Nets}

\author{Martin Schaller}
\email{martin_schaller@acm.org}
\affiliation{Algorithmics,
Parkring 10, 1010  Vienna, Austria}

\author{Karl Svozil}
\email{svozil@tuwien.ac.at}
\homepage{http://tph.tuwien.ac.at/~svozil}
\affiliation{Institut f\"ur Theoretische Physik, University of Technology Vienna,
Wiedner Hauptstra\ss e 8-10/136, A-1040 Vienna, Austria}

\begin{abstract}
Two novel computing models based on an infinite tessellation of space-time are introduced. They consist of recursively coupled primitive building blocks. The first model is a scale-invariant generalization of cellular automata, whereas the second one utilizes self-similar Petri nets. Both models are capable of hypercomputations and can, for instance, ``solve'' the halting problem for Turing machines. These two models are closely related, as they exhibit a step-by-step equivalence for finite computations. On the other hand, they differ greatly for computations that involve an infinite number of building blocks: the first one shows indeterministic behavior whereas the second one halts. Both models are capable of challenging our understanding of computability, causality, and space-time.
\end{abstract}


\maketitle

\section{Introduction}


Every physically relevant computational model must be mapped into physical space-time and {\it vice versa}
\cite{landauer-89,maxwell-demon,bennett-73}.
In this line of thought, Von Neumann's self-reproducing Cellular Automata \cite{v-neumann-66}
have been envisioned by Zuse \cite{zuse-69}
and other researchers  \cite{fredkin,toffoli-margolus-90,wolfram-2002,Margenstern:jucs_5_9:a_polynomial_solution_for}
as  ``calculating space;''
i.e., as a  locally connected grid of finite automata~\cite{hopcroft}
capable of universal algorithmic tasks, in which
intrinsic~\cite{svozil-94} observers are embedded~\cite{toffoli:79}.
This model is conceptually discreet and
noncontinuous and resolves the eleatic ``arrow''
antinomy \cite{zeno,ki-57,gruenbaum:68,salmon-01}
against motion in discrete space by introducing
the concept of information about the state of motion in between time steps.

Alas,  there is no direct physical evidence supporting  the assumption of a tessellation of configuration space or time.
Given enough energy, and without the possible bound at the Planck length of about $10^{-35}$m, physical configuration space seems
to be potentially infinitely divisible.
Indeed, infinite divisibility of space-time has been utilized for proposals of a kind of ``Zeno oracle''~\cite{weyl:49},
a progressively accelerated Turing machine~\cite{gruenbaum:74,Davies01,ord-2006}
capable of hypercomputation~\cite{Davis-2004,Doria-2006,Davis-2006}.
Such accelerated Turing machines have also been discussed in the relativistic context~\cite{DBLP:conf/mcu/Durand-Lose04,Nemeti2006118}.
In general, a physical model capable of hypercomputation by some sort of ``Zeno squeezing'' has to cope with two seemingly contradictory
features: on the one hand, its infinite capacities could be seen as an obstacle of evolution and therefore
require a careful analysis of the principal possibility of motion in finite space and time
{\it via} an infinity of cycles or stages.
On the other hand, the same infinite capacities could be perceived as an advantage, which
might yield algorithms beyond the Turing bound of universal computation, thus extending the Church-Turing thesis.

The models presented in this article unify the connectional clarity of
von Neumann's Cellular Automaton model with the requirement of infinite divisibility of cell space.
Informally speaking,
the scale-invariant cellular automata presented ``contain''  a multitude of ``spatially'' and ``temporally'' ever decreasing copies of themselves,
thereby using different time scales at different layers of cells.
The cells at different levels are also capable to communicate, i.e., exchange information, with these copies, resulting in ever smaller and faster cycling cells.
The second model is based on Petri nets which can enlarge themselves.

The advantage over existing models of accelerated Turing machines
--- which are just Turing machines with a geometrically progression of accelerated time cycles ---
resides in the fact that the underlying computational medium is embedded into its environment in a uniform and homogeneous way.
In these new models, the entire universe, and not just specially localized parts therein, is uniformly capable of the same computational capacities.
This uniformity of the computational environment could be perceived as one further step towards the formalization of continuous physical systems~\cite{CIEChapter2007} in algorithmic terms.
In this respects, the models seem to be closely related to classical continuum models, which are at least in principle capable of
unlimited divisibility and information flows at arbitrary small space and time dimensions.
At present however, for all practical purposes, there are finite bounds on divisibility and information flow.

To obtain a taste of some of the issues encountered in formalizing this approach, note that
an infinite sequence of ever smaller and faster cycling cells leads to the following situation.
Informally speaking, let a {\em self-similar cellular automaton} be a variant of a one-dimensional elementary cellular automaton,
such that each cell is updated twice as often as its left neighbor.
The cells of a self-similar cellular automaton can be enumerated as $\ldots, c_{-2}, c_{-1}, c_{0}, c_{1}, c_{2}, \ldots$.
Starting at time 0 and choosing an appropriate time unit, cell $c_i$ is updated at
times $1 / 2^{i}, 2 /2^{i}, 3 / 2^{i}, \ldots$.
Remarkably, this definition leads to indeterminism.
To see this, let $s(i, t)$ be the state of cell $i$ at time $t$.
Now, the state $s(0, 1)$ depends on $s(1, 1/2)$, which itself depends on $s(2, 1/2^2)$ and so on,
leading to an infinite regress.
In general, in analogy to Thomson's paradox~\cite{salmon-01,1011191}, this results in an undefined or at least nonunique and thus indeterministic behavior of the automaton.

This fact relates to the following variant of Zeno's paradox of a runner,
according to which the runner cannot even get started~\cite{salmon-01}.
He must first run to the half way point, but before that he must run half way
to the half way point and so on indefinitely.
Whereas Zeno's runner can find rescue in the limit of convergent real sequences,
there is no such relieve for the discrete systems considered.

Later on, two restrictions on self-similar automata (build from scale-invariant cellular automata) are presented,
which are sufficient conditions for deterministic behavior, at least for finite computations.
Furthermore, a similar model based on a variant of Petri nets will be introduced, that avoids indeterminism and
halts in the infinite limit, thereby coming close to the spirit of Zeno's paradox.


The article is organized as follows.
Section \ref{sec-tm} defines the  Turing machine model used in the remainder of the article,
and introduces two hypercomputing models: the accelerated and the right-accelerated Turing machine.
In section \ref{chap:sica} self-similar as well as scale-invariant cellular automata are presented.
Section \ref{chap:hypercomputer} is devoted to the construction of a hypercomputer based on
self-similar cellular automata.
There is a strong resemblance between this construction and the right-accelerated Turing machine,
as defined in section \ref{sec-tm}.
A new computing model, the self-similar Petri net  is introduced in section \ref{chap:petri}.
This model features a step-to-step equivalence to self-similar cellular automata for finite computations, but halts
in the infinite case.
The same construction as in section \ref{chap:hypercomputer} is used to demonstrate that self-similar Petri nets
are capable of hypercomputation.
The final section contains some concluding remarks and gives some directions for future research.

\section{Turing machines and accelerated Turing machines}
\label{sec-tm}

The Turing machine is, beside other formal systems that are computationally equivalent, the most powerful model
of classical computing~\cite{rogers1,odi:89,odi:99}.
We use the following model of a Turing machine \cite{hopcroft}.

\begin{defn}[Turing Machine]
Formally, a {\em Turing machine}  is a tuple
$M = (Q, \Sigma, \Gamma, \delta, q_0, B, F)$,
where $Q$ is the finite set of states, $\Gamma$ is the finite set of tape symbols,
$\Sigma \subset \Gamma$ is the set of input symbols, $q_0 \in Q$ is the start state,
$B \in \Gamma \backslash \Sigma$ is the blank, and $F \subset Q$ is the set of final states.
The next move function or transition function $\delta$ is a mapping from
$Q \times \Gamma$ to $Q \times \Gamma \times \{L, R\}$, which may be undefined for some arguments.
\end{defn}

The Turing machine $M$ works on a tape divided into cells that has a leftmost cell but is infinite to the right.
Let $\delta(q, a) = (p, b, D)$.
One step (or move) of $M$ in state $q$ and the head of $M$ positioned over input symbol $a$
consists of the following actions:
scanning input symbol $a$, replacing symbol $a$ by $b$,
entering state $p$ and moving the head one cell either to the left ($D=L$) or to the right ($D=R$).
In the beginning, $M$ starts in state $q_0$ with a tape that is initialized with an input word $w \in \Sigma^*$,
starting at the leftmost cell, all other cells blank,
and the head of $M$ positioned over the first symbol of $w$.
We need sometimes the function $\delta$ split up into three separate functions:
$\delta(q,a) = (\delta_Q(q,a), \delta_\Gamma(q,a), \delta_D(q,a))$.

The configuration of a Turing machine $M$ is denoted by a string of the form
$\alpha_1 q \alpha_2$, where $q \in Q$ and $\alpha_1, \alpha_2 \in \Gamma^*$.
Here $q$ is the current state of $M$, $\alpha_1$ is the tape content to the left,
and $\alpha_2$ the tape content to the right of the head including the symbol that is scanned next.
Leading and trailing blanks will be omitted, except the head has moved to the left or to the right of
the non-blank content.
Let $\alpha_1 q \alpha_2$ and  $\alpha_1^\prime p \alpha_2^\prime$ be two configurations of $M$.
The relation $\alpha_1 q \alpha_2 \vdash_M \alpha_1^\prime p \alpha_2^\prime$ states
that $M$ with configuration $\alpha_1 q \alpha_2$ changes in one step
to the configuration $\alpha_1^\prime p \alpha_2^\prime$.
The relation $\vdash_M^*$ denotes the reflexive and transitive closure of $\vdash_M$.

The original model of a Turing machine as introduced by Alan Turing contained no statement about the time in which a
step of the Turing machine has to be performed.
In classical computation, a "yes/no"-problem is therefore decidable if, for each problem instance, the answer is obtained
in a finite number of steps.
Choosing an appropriate time scheduling, the Turing machine can perform infinitely many steps in finite time, which
transcends classical computing,
thereby leading to the following two hypercomputing models.
The concept of an accelerated Turing machine was independently proposed by Bertrand Russell, Ralph Blake,
Hermann Weyl and others (see Refs.~\cite{ord-2006,potgieter-06}).

\begin{defn}[Accelerated Turing machine]
An accelerated Turing machine is a Turing machine which performs the $n$-th step of a calculation in $1/2^n$ units of time.
\end{defn}

The first step is performed in time 1, and each subsequent step in half of the time before.
Since $1 + 1/2 + 1/4 + 1/8 + \ldots = 2$, the accelerated Turing machine can perform infinitely many steps
in finite time.
The accelerated Turing machine is a hypercomputer, since it can, for example, solve the halting problem, see e.g., Ref.~\cite{ord-2006}.
If the output operations are not carefully chosen, the state of a cell becomes indeterminate, leading to
 a variation of Thomson's lamp paradox.
The open question of the physical dynamics in the limit
reduces the physical plausibility of the model.
The following model of a hypercomputing Turing machine has a different time scheduling, thereby avoiding
some of the paradoxes that might arise from the previous one.

\begin{defn}[Right-accelerated Turing machine]
Let the cells of the tape be numbered from the left to the right $c_0, c_1, c_2, \ldots$.
A right-accelerated Turing machine is a Turing machine that takes $1/2^n$ units of time to perform a step that moves the head from cell
$c_n$ to one of its neighbor cells.
\end{defn}

\begin{theorem}
\label{th-right-acc-tm}
There exists a right-accelerated Turing machine that is a hypercomputer.
\end{theorem}

\begin{proof}
Let $M_U$ be a universal Turing machine.
We construct a Turing machine $\overline{M}_U$ that alternates between simulating one step of $M_U$ and shifting over
the tape content one cell to the right.
We give a sketch of the construction, Ref.~\cite{hopcroft} contains a detailed description of the used techniques.
The tape of $\overline{M}_U$ contains one additional track that is used
to mark the cell  that is read next by the simulated $M_U$.
The finite control of $\overline{M}_U$ is able to store simultaneously the state of the head of $M_U$ as well
as a tape symbol of $M_U$.
We assume that the input of $M_U$ is surrounded by two special tape symbols, say $\$$.
At the start of a cycle, the head of $\overline{M}_U$ is initially positioned over the left delimiter $\$$.
$\overline{M}_U$ scans the tape to the right, till it encounters a flag in the additional track that marks the head position of $M_U$.
Accessing the stored state of $M_U$, $\overline{M}_U$ simulates one step of $M_U$ thereby marking
either the left or the right neighbor cell as the cell that has to be visited next in the simulation of $M_U$.
If necessary, a blank is inserted left to the right delimiter $\$$, thereby extending the simulated tape of $M_U$.
Afterwards the head of $\overline{M}_U$ moves to the right delimiter $\$$ to start the shift over that is
performed from the right to the left.
$\overline{M}_U$ repeatedly stores the symbols read in its finite control and prints them to the cell to the right.
After the shift over, the head of $M_U$ is positioned over the left delimiter $\$$ which finishes one cycle.

We now give an upper bound of the cycle time.
Let $n$ be the number of cells, from the first $\$$ to the second one.
Without loss of generality we assume that $c_0$ contains the left $\$$.
$\overline{M}_U$ scans from the left to the right and simulates one step of $M_U$ which might require to go an
additional step to the left.
If cell $c_1$ is to be read next, the head of $M_U$ cannot move to the right, otherwise it would fall off the
tape of $M_U$.
Therefore the worst case occurs if the cell $c_2$ is marked as cell that $M_U$ has to be read next.
In this case we obtain $1 + 1/2 + 1/4 + 1/2 + 1/4 + 1/8 + \ldots  + 1/2^{n-1} < 3$.
The head of $\overline{M}_U$ is now either over cell $c_{n-1}$, or over cell $c_n$ if a insertion was performed.
The shift over visits each cell $c_i, 1 \leq i < n$ three times, and $c_0$ two times.
Therefore the following upper bound of the time of the shift over holds:
$3(1 + 1/2 + 1/4 + \ldots 1/2^n) < 6$.
We conclude that if the cycle started initally in cell $c_n$ it took less than time $9/2^n$.
If $M_U$ halts on its input, $\overline{M}_U$ finishes the simulation in a time less than
$9(1 + 1/2 + 1/4 + \ldots) = 18$.
$\overline{M}_U$ therefore solves the halting problem of Turing machines.
We remark that if $M_U$ does not halt, the head of $\overline{M}_U$ vanishes in infinity, leaving a blank tape behind.
\end{proof}

A right-accelerated Turing machine is, in contrast to the accelerated one, in control over
the acceleration.
This can be used to transfer the result of a computation back to slower cells.
The construction of an infinite machine, as proposed by Davies~\cite{Davies01}, comes close to the model of a right-accelerated Turing machine,
and his reasoning shows that a right-accelerated Turing machine could be build within a continuous Newtonian universe.

\section{Self-similar and scale-invariant cellular automata}
\label{chap:sica}

\subsection{Basic definitions}

\emph{Cellular automata} are dynamical systems in which space and time are discreet.
The states of cells in a regular lattice are updated synchronously according to a local deterministic
interaction rule.
The rule gives the new state of each cell as a function of the old states of some ``nearby'' neighbor cells.
Each cell obeys the same rule, and has a finite (usually small) number of states.
For a more comprehensive introduction to cellular automata, we refer to Refs.~\cite{v-neumann-66,wolfram-86,gutowitz,ilachinski01,wolfram-2002}.

A {\em scale-invariant cellular automaton}  operates like an ordinary
{\em cellular automaton}  on a cellular space, consisting of a regular arrangement of cells,
whereby each cell can hold a value from a finite set of states.
Whereas the cellular space of a cellular automaton consists of a regular one- or higher
dimensional lattice, a scale-invariant cellular automaton operates on a cellular space of recursively nested lattices
which can be embedded in some Euclidean space as well.

The time behavior of a scale-invariant cellular automaton differs from the time behavior of a cellular automaton:
Cells in the same lattice synchronously change their state \cite{Morelli_Zanette}, but
as cells are getting smaller in deeper nested lattices, the time steps between state changes in
the same lattice are assumed to {\em decrease} and approach zero in the limit.
Thereby, a finite speed of signal propagation between adjacent cells is always maintained.
The scale-invariant cellular automaton model gains its  computing capabilities by introducing a local rule that
allows for interaction between adjacent lattices \cite{BoFeng_MengDing}.
We will introduce the scale-invariant cellular automaton model for the one-dimensional case, the extension to higher dimensions
\cite{Brunnet_Chate} is
straightforward.

\begin{figure}
\begin{center}
\scalebox{0.7}{\includegraphics{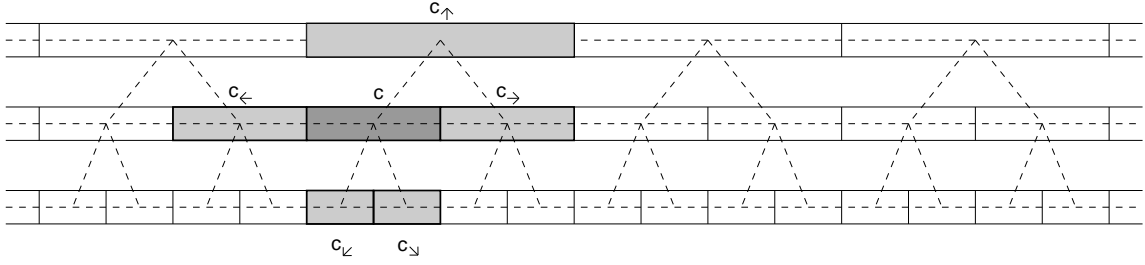}}
\caption{\label{fig:1-dim-interaction} Space and topological structure of a scale-invariant cellular automaton.}
\end{center}
\end{figure}

A scale-invariant cellular automaton, like a cellular automaton, is defined by a cellular space, a topology that defines the neighborhood of a cell, a finite set of
states a cell can be in,  a time model that determines when a cell is updated, and a local rule that
maps states of neighborhood cells to a state.
We first define the cellular space of a scale-invariant cellular automaton.
To this end, we make use of standard interval arithmetic.
For a scalar $\lambda \in \mathbb{R}$ and a (half-open) interval $[x,y) \subset \mathbb{R}$ set:
$\lambda + [x,y) = [\lambda + x, \lambda + y)$ and $\lambda [x,y) = [\lambda x, \lambda y)$.
We denote the unit interval $[0,1)$ by $\mathbbm{1}$.

\begin{defn}[Cellular Space and Space Operators]
The cellular space $\mathcal{C}$, the set of all cells of the scale-invariant cellular automaton,
is the set $\mathcal{C} = \{ 2^k (i + \mathbbm{1}) | i, k \in \mathbb Z\}$.
The neighborhood of a cell $c$ is determined by the following operators $\mathit{op}: \mathcal{C} \rightarrow \mathcal{C}$.
For a cell $c = 2^k (i + \mathbbm{1})$ in $\mathcal{C}$ let
$c_{\leftarrow} = 2^k (i - 1 + \mathbbm{1})$ be the left neighbor,
$c_{\rightarrow} = 2^k (i + 1 + \mathbbm{1})$ the right neighbor,
$c_{\uparrow} =  2^{k + 1} (\lfloor \frac{i}{2} \rfloor + \mathbbm{1})$ the parent,
$c_{\swarrow} = 2^{k-1}(2i + \mathbbm{1})$ the left child,
and $c_{\searrow} = 2^{k-1}(2i + 1 + \mathbbm{1})$ the right child of $c$.
The predicate $\mathit{left}(c)$ is true if and only if the cell $c$ is the left child of its parent.
\end{defn}
The cellular space $\mathcal{C}$ is the union of all lattices $L_k=\{2^k (i + \mathbbm{1})| i \in \mathbb Z\}$, where $k$ is an integer.
This topology is depicted in Fig.~\ref{fig:1-dim-interaction}.
For notational convenience, we introduce a further operator, this time from $\mathcal{C}$ to $\mathcal{C} \times \mathcal{C}$,
that maps a cell to its both child cells:
 $c_{\downarrow} = (c_{\swarrow}, c_{\searrow})$.
We remark that according to the last definition for each cell either $\mathit{left}(c)$ or $\neg \mathit{left}(c)$ is true.
Later on, we will consider scale-invariant cellular automata where not each cell has a parent cell.
If $c = 2^k (i + \mathbbm{1})$ is such a cell, we set by convention $\mathit{left}(c) = 1$ if $i \mod 2 = 0$, otherwise $\mathit{left}(c) = 0$.

\begin{figure}
\begin{center}
\scalebox{0.7}{\includegraphics{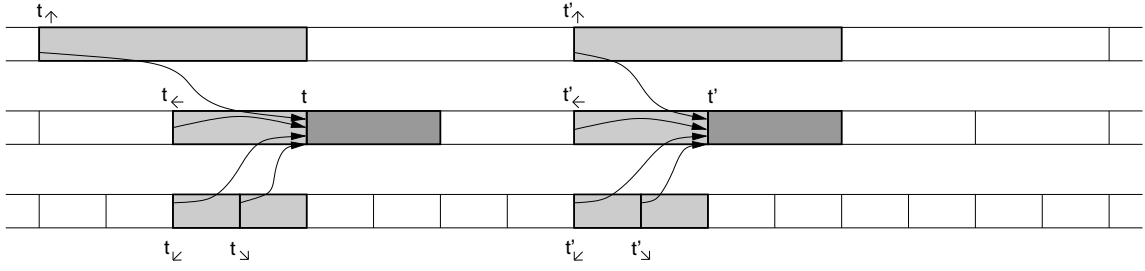}}
\caption{\label{fig:timeops} Temporal dependencies of a scale-invariant cellular automaton.}
\end{center}
\end{figure}

All cells in lattice $L_k$ are updated synchronously at time instances $2^k i$ where $i$ is an integer.
The time interval between two cell updates in lattice $L_k$ is again a half-open interval $2^k (i + \mathbbm{1})$
and the cycle time, that is the time between two updates of the cell, is therefore $2^k$.
A simple consequence of this time model is that child cells cycle twice as fast and the parent cell cycle
half as fast as the cell itself.
\begin{defn}[Time Scale and Time Operators]
The time scale $\mathcal{T}$ is the set of all possible time intervals, which is in the one-dimensional
case equal to the set $\mathcal{C}$: $\mathcal{T} = \{ 2^k (i + \mathbbm{1}) | i, k \in \mathbb Z\}$.
The temporal dependencies of a cell are expressed by the following time operators $\mathit{op}: \mathcal{T} \rightarrow \mathcal{T}$.
For a time inverval $t = 2^k (i + \mathbbm{1})$ let
$t_\leftarrow = 2^k (i - 1 + \mathbbm{1})$,
$t_\uparrow = 2^{k + 1} (\lfloor \frac{i-1}{2} \rfloor + \mathbbm{1})$,
$t_\swarrow = 2^{k-1} (2i - 2  + \mathbbm{1})$, and
$t_\searrow =  2^{k-1} (2i - 1 + \mathbbm{1})$.
The predicate $\mathit{coupled}(t)$ is true if and only if the state change of a cell at the beginning of $t$ occurs
simultaneously with the state change of its parent cell.
\end{defn}
The usage of time intervals instead of time instances, has the advantage
that a time interval uniquely identifies the lattice where the update occurs.
Fig.~\ref{fig:timeops} depicts the temporal dependencies of a cell: to the left it shows a
coupled state change, to the right an uncoupled one.
We remark that we denoted space and time operators by the same symbols, even if their mapping is different.
In applying these operators, we take in the remainder of this paper care, that the context of the operator is always clearly defined.

At any time, each cell is in one state from a finite state set $Z$.
The cell state in a given time interval is described by the state function $s(c,t)$,
which maps cells and time intervals to the state set.
The space-time scale $\mathcal{S}$ of the scale-invariant cellular automaton describes the set of allowed pairs of cells and time intervals:
$\mathcal{S} = \{(c,t)| c \in \mathcal{C}, t \in \mathcal{T} \mbox{ and } |c| =|t|\}$.
Then, the state function $s$ can be expressed as a mapping $s: \mathcal{S} \rightarrow Z$.
The local rule describes the evolution of the state function.
\begin{defn}[Local Rule]
For a cell $c$ and a time interval $t$, where $(c, t)$ is in $\mathcal{S}$, the
evolution of the state is given by the local rule $f$ of the scale-invariant cellular automaton
\begin{equation}
\label{eq:local-rule}
\small
s(c,t) = f(
        s(c_\uparrow, t_\uparrow),
  s(c_\leftarrow, t_\leftarrow), s(c, t_\leftarrow), s(c_\rightarrow, t_\leftarrow),
  s(c_\downarrow, t_\swarrow),  s(c_\downarrow, t_\searrow), \mathit{left}(c), \mathit{coupled}(t)
)
\end{equation}
\end{defn}
In accordance with the definition, the expanded form of a expression of the kind $s(c_\downarrow, t_\swarrow)$ is
 $(s(c_\swarrow, t_\swarrow), s(c_\searrow, t_\swarrow))$.
The local rule $f$ is a mapping from $Z^8 \times \{0,1\}^2$ to $Z$.
Beside the dependencies on the states of the neighbor cells, the new state of the cell further depends on
whether the cell is the left or the right child of its parent cell and
whether the state change is coupled or uncoupled to the state change of its parent cell.
Formally, a scale-invariant cellular automaton $A$ is denoted by the tuple $A = (Z, f)$.
There are some simplifications of the local rule possible, if one allows for a larger state set.
For instance, the values of the predicates $\mathit{left}$ and $\mathit{coupled}$ could be stored as
substate in the initial configuration.
If the local rule accordingly updates the value of $\mathit{coupled}$, the dependencies on the boolean
predicates could be dropped from the local rule.

As noted in the introduction the application of the local rule in its general form might lead to indeterministic behavior.
The next subsection introduces two restrictions of the general model that avoid indeterminism at least for finite computations.
A special case of the local rule is a rule of the form
$f(s(c_\leftarrow, t_\leftarrow), s(c, t_\leftarrow), s(c_\rightarrow, t_\leftarrow))$,
which is the constituting rule of a one-dimensional 3-neighborhood cellular automaton.
In this case, the scale-invariant cellular automaton splits up in a sequence of infinitely many nonconnected cellular automata.
This shows that the scale-invariant cellular automaton model is truly an extension of the cellular automaton model and allows us to
view a scale-invariant cellular automaton as an infinite sequence of interconnected cellular automata.

We now examine the signal speed that is required to communicate state changes between neighbor cells.
To this end, we select the middle point of a cell as the source and the target of a
signal that propagates the state change of a cell to one of its neighbor cells.
A simple consideration shows that the most restricting cases are the paths from the space time points
$(c_\leftarrow, t_\leftarrow)$, $(c_\uparrow, t_\uparrow)$, $(c_\swarrow, t_\searrow)$ to $(c,t)$ if not $\mathit{coupled}(t)$.
The simple calculation delivers the results $1,1$, and $\frac{1}{2}$, respectively,
hence a signal speed of 1 is sufficient to deliver the updates in the given timeframe.
A more general examination takes also the processing time of a cell into account.
If a cell in $L_k$ takes time $2^k p$ to process their inputs and if we assume a finite signal speed of $v$,
the cycle time of a cell in $L_k$ must be at least $2^k (p + v)$.
In sum, as long as the processing time is proportional to the diameter of a cell,
we can always find a scaling factor $t \rightarrow \lambda t$, such that
the scale-invariant cellular automaton has cycle times that conform to the time scale $\mathcal{T}$.

\subsection{Self-similar cellular automata and indeterminism}

The construction of a hypercomputer in section \ref{chap:hypercomputer} uses a simplified version of a scale-invariant cellular automaton, which we call
a Self-similar Cellular Automaton.
\begin{defn}[Self-similar Cellular Automaton]
A {\em self-similar cellular automaton} has the cellular space $\mathcal{C} = \{2^k \mathbbm{1} | k \in \mathbb Z\}$,
the time scale $\mathcal{T} = \{2^k (i + \mathbbm{1})| i, k \in \mathbb{Z}\}$, and the finite state set $Z$.
The space-time scale of a self-similar cellular automaton is the set $\mathcal{S} = \{(c,t)| c \in \mathcal{C}, t \in \mathcal{T} \mbox{ and } |c| = |t|\}$.
The self-similar cellular automaton has the following local rule: for all $(c, t) \in \mathcal{S}$
\begin{equation}
s(c,t) =
f(
        s(c_\uparrow, t_\uparrow),
        s(c, t_\leftarrow),
        s(c_\swarrow, t_\swarrow),
        s(c_\swarrow, t_\searrow),
        \mathit{coupled}(t)
)
\end{equation}
\end{defn}

The local rule $f$ is a mapping from $Z^4 \times \{0,1\}$ to $Z$.
Formally, a self-similar cellular automaton $A$ is denoted by a tuple $A = (Z, f)$.
By restricting the local rule of a scale-invariant cellular automaton, a self-similar cellular automaton can also be constructed from a scale-invariant cellular automaton.
Consider a scale-invariant cellular automaton whose local rule does not depend on
the cell neighbors $c_\leftarrow$, $c_\rightarrow$, and $c_\searrow$.
Then, the resulting scale-invariant cellular automaton contains the self-similar cellular automaton as subautomaton.

We introduce the following notation for self-similar cellular automata.
We index a cell $[0, 2^k)$ by the integer $-k$, that is a cell with index $k$ has a cyle time of $2^{-k}$.
We call the cell $k - 1$ the upper neighbor and the cell $k + 1$ the lower neighbor of cell $k$.
Time instances can be conveniently expressed as a binary number.
If not stated otherwise, we use the cycle time of cell 0 as time unit.

We noted already in the introduction that the evolution of a scale-invariant cellular automaton might lead to indeterministic behavior.
We offer two solutions, one based on a special quiescent state, the other one based on a dynamically growing lattice.

\begin{defn}[Short-Circuit Evaluation]
A state $q$ in the state set $Z$ is called a quiescent state with regard to the short-circuit evaluation, if
$f(q, q, ?, ?, ?) =  q$, where the question mark sign ``$?$'' either represents an arbitrary state or a boolean value, depending on
its position.
Whenever a cell is in state $q$, the cell does not access its lower neighbor.
\end{defn}

The cell remains as long in the quiescent state as long as the upper neighbor is in the quiescent state, too.
This modus of operandi corresponds to the short-circuit evaluation of logical expressions in programming
languages like C or Java.
If the self-similar cellular automaton starts in an initial configuration of the form $z_0 z_1 \ldots z_n q q q\ldots$ at cell $0$, the infinite
regress is interrupted, since cell $n+2$ evolves to $q$ without being dependent on cell $n+3$.

\begin{defn}[Dynamically growing self-similar cellular automaton]
Let $q$ be a state in the state set $Z$, called the quiescent state.
A dynamically growing self-similar cellular automaton initially starts with the finite set of cells $0, \ldots, n$ and the following
boundary condition.
Whenever cell $0$ or the cell with the highest index $k$ is evolved, the state of the missing neighbor cell is assumed to be $q$.
The self-similar cellular automaton dynamically appends cells to the lower end when needed:
whenever the cell with the highest index $k$ enters a state that is different from the quiescent state,
a new cell $k + 1$ is appended, initialized with state $q$, and connected to the cell $k$.
To be more specific: If $k$ is the highest index, and cell $k$ evolves at time $2^{-k} i$ to state $z \neq q$,
a new cell $k + 1$ in state $q$ is appended.
The cell performs its first transition at time $2^{-k}(i + 1/2)$, assuming state $q$ for its missing
lower neighbor cell.
\end{defn}

We note that the same technique could also be applied to append upper cells to the self-similar cellular automaton, although in the remainder of this paper we only deal
with self-similar cellular automata which are growing to the bottom.
Both enhancements ensure a deterministic evaluation either for a configuration where only a finite number of cells is in a nonquiescent state
or for a finite number of cells.

\begin{defn}[Finite and Final Configuration]
A configuration of a self-similar cellular automaton $A$ is called finite if only a finite number of cells is different from the quiescent state.
Let $C$ be a finite configuration and $C^\prime$ the next configuration in the evolution that is
different to $C$.
$C^\prime$ is again finite.
We denote this relationship by $C \vdash_{A} C^\prime$.
The relation $\vdash_{A}^*$ is again the reflexive and transitive closure of $\vdash_{A}$.

A self-similar cellular automaton as a scale-invariant cellular automaton cannot halt by definition  and runs forever without stopping.
The closest analogue to the Turing machine halting occurs, when the configuration stays constant during evolution.
Such a configuration that does not change anymore is called final.
\end{defn}

\section{Constructing a hypercomputer}
\label{chap:hypercomputer}
In this section, we shall construct an accelerated Turing machine based on a self-similar cellular automaton.
A self-similar cellular automaton which simulates the Turing machine $\overline{M}_U$ specified in the  proof of Theorem~\ref{th-right-acc-tm} in a
step-by-step manner is a hypercomputer, since the resulting Turing machine is a right-accelerated one.
We give an alternative construction, where the shift over to the right is directly embedded in the local rule of the self-similar cellular automaton.
The self-similar cellular automaton will simultaneously simulate the Turing machine and shift the tape content down
to faster cycling cells.
The advantages of this construction are the smaller state set as well as a resulting faster simulation.

\subsection{Specification}

Let $M = (Q, \Sigma, \Gamma, \delta, q_0, B, F)$ be an arbitrary Turing machine.
We construct a self-similar cellular automaton $A_M = (Z, f)$ that simulates $M$ as follows.
First, we simplify the local rule by dropping the dependency on $t_\swarrow$, obtaining
\begin{equation}
s(c,t) =
f(
        s(c_\uparrow, t_\uparrow),
        s(c, t_\leftarrow),
        s(c_\swarrow, t_\searrow),
        \mathit{coupled}(t)
).
\end{equation}
The state set $Z$ of $A_M$ is given by
\[
Z = \Gamma \cup (\Gamma \times \{\rightarrow\}) \cup (Q \times \Gamma)
\cup (Q \times \Gamma \times \{\rightarrow\}) \cup
\{\Box, \blacktriangleleft, \lhd, \overrightarrow{\lhd}, \rhd, \rhd_B, \rhd_\blacktriangleleft\}.
\]
We write $\overrightarrow{a}$ for an element $(a, \rightarrow)$ in $\Gamma \times \{\rightarrow\}$,
$\langle q,a \rangle$ for an element
$(q, a)$ in $Q \times \Gamma$, and
$\overrightarrow{\langle q,a \rangle}$ for an element
$(q, a, \rightarrow)$ in $Q \times \Gamma \times \{\rightarrow\}$.
To simulate $M$ on the input $w=a_1 \ldots a_n$ in $\Sigma^*$, $n \geq 1$,
$A_M$ is initialized with the sequence
$\overrightarrow{\lhd} \langle q_0,a_1 \rangle a_2 a_3\ldots a_n\rhd $
starting at cell 0, all other cells shall be in the quiescent state $\Box$.
If $w=a_1$, $A_M$ is initialized with the sequence
$\overrightarrow{\lhd} \langle q_0,a_1 \rangle B\rhd $, and
if $w=\epsilon$, the empty word, $A_M$ is initialized with the sequence
$\overrightarrow{\lhd} \langle q_0,B \rangle B\rhd $.
We denote the initial configuration by $C_0$, or by $C_0(w)$ if we want to emphasize the dependency on the input word $w$.
The computation is started at time 0, i.e. the first state change of cell $k$ occurs at time $2^{-k}$.

The elements $\langle q, a \rangle$ and $\overrightarrow{\langle q, a \rangle}$  act as head of the
Turing machine including the input symbol of the Turing machine that is scanned next.
To accelerate the Turing machine, we have to shift down the tape content to faster cycling cells of the self-similar cellular automaton,
thereby taking care that the symbols that represent the non-blank content of the Turing machine tape are kept together.
We achieve this by sending a pulse, which is just a symbol from a subset of the state set,
from the left delimiter $\lhd$ to the right delimiter $\rhd$ and back.
Each zigzag of the pulse moves the tape content one cell downwards and triggers
at least one move of the Turing machine.
Furthermore a blank is inserted to the right of the simulated head if necessary.
The pulse that goes down is represented by exactly one element of the form
$\overrightarrow{\lhd}, \overrightarrow{a}, \overrightarrow{\langle q,a \rangle}, \rhd_B$, or $\rhd_\blacktriangleleft$,
the upgoing pulse is represented by the element $\blacktriangleleft$.

The specification of the values for the local rule $f$ for all possible arguments
is tedious, therefore we use the following approach.
A coupled transition of two neighbor cells can perform a simultaneous state change of the two cells.
If the state changes of these two neighbor cells is independent of their other neighbors,
we can specify the state changes as a transformation of a state pair into another one.
Let $z_1, z_2, z_1^\prime, z_2^\prime$ be elements in $Z$.
We call a mapping of the form $z_1 \: z_2 \mapsto z_1^\prime \: z_2^\prime$ a block transformation.
The block transformation $z_1 \: z_2 \mapsto z_1^\prime \: z_2^\prime$ defines
 a function mapping of the form
$
f(x, z_1, z_2, 0) = f(x, z_1, z_2, 1) = z_1^\prime
$
and
$
f(z_1, z_2,y, 1) = z_2^\prime
$
for all $x, y$ in $Z$.
Furthermore, we will also allow block transformations that might be ambiguous for certain configurations.
Consider the block transformations
$z_1 \: z_2 \mapsto z_1^\prime \: z_2^\prime$
and
$z_2 \: z_3 \mapsto z_2^{\prime\prime} \: z_3^\prime$
that might lead to an ambiguity for a configuration that contains $z_1z_2z_3$.
Instead of resolving these ambiguities in a formal way, we will restrict our consideration to
configurations that are unambiguous.

The evolution of the self-similar cellular automaton $A_M$ is governed by the following block transformations:
\begin{enumerate}

\item
\emph{Pulse moves downwards.}
Set
\begin{equation}
\overrightarrow{\lhd} \: \langle q, a \rangle \mapsto \lhd \:
\overrightarrow{\langle q, a \rangle};
\label{tr:start-state}
\end{equation}
\begin{equation}
\overrightarrow{a} \: b \mapsto a \: \overrightarrow{b};
\label{tr:down}
\end{equation}
\begin{equation}
\overrightarrow{\lhd} \:a \mapsto \lhd \: \overrightarrow{a}.
\label{tr:start}
\end{equation}
If $\delta(q,a) = (p,c,R)$ set
\begin{equation}
\overrightarrow{b} \: \langle q, a \rangle \mapsto b \:
\overrightarrow{\langle q, a \rangle};
\label{tr:down-to-head}
\end{equation}
\begin{equation}
\overrightarrow{\langle q,a \rangle} \: b \mapsto c \:
\overrightarrow{\langle p, b \rangle};
\label{tr:right-2}
\end{equation}
\begin{equation}
\overrightarrow{\langle q,a \rangle} \: \rhd \mapsto \langle q,a \rangle \:
\rhd_B.
\label{tr:down-state-right-delimiter-blank}
\end{equation}
If $\delta(q,a) = (p,c,L)$ set
\begin{equation}
\overrightarrow{b} \: \langle q, a \rangle \mapsto \langle p, b \rangle \:
\overrightarrow{c};
\label{tr:left-1}
\end{equation}
\begin{equation}
\overrightarrow{\langle q,a \rangle} \: b \mapsto \langle q,a \rangle \:
\overrightarrow{b};
\label{tr:left-no-move}
\end{equation}
\begin{equation}
\overrightarrow{\langle q,a \rangle} \: \rhd \mapsto \langle q,a \rangle \:
\rhd_\blacktriangleleft.
\label{tr:down-state-right-delimiter}
\end{equation}
Set
\begin{equation}
\overrightarrow{a} \: \rhd \mapsto a \: \rhd_\blacktriangleleft;
\label{tr:down-a-rhd}
\end{equation}
\begin{equation}
\rhd_B \: \Box \mapsto B \: \rhd_\blacktriangleleft;
\label{tr:new-blank}
\end{equation}
\begin{equation}
\rhd_\blacktriangleleft \: \Box \mapsto \blacktriangleleft \: \rhd.
\label{tr:reflection-right}
\end{equation}

\item
\emph{Pulse moves upwards}.
Set
\begin{equation}
a \: \blacktriangleleft \mapsto \blacktriangleleft \: a;
\label{tr:up}
\end{equation}
\begin{equation}
\langle q,a \rangle \: \blacktriangleleft \mapsto  \blacktriangleleft \:
\langle q,a \rangle;
\label{tr:up-state}
\end{equation}
\begin{equation}
\lhd \: \blacktriangleleft \mapsto \Box \: \overrightarrow{\lhd}.
\label{tr:up-lhd}
\end{equation}
\end{enumerate}

If to a certain cell no block transformation is applicable the cell shall remain in its previous state.
Furthermore, we assume a short-circuit evaluation with regard to the quiescent state:
$f(\Box, \Box, ?, ?) = \Box$, whereby the lower neighbor cell is not accessed.

\subsection{Example}

\begin{figure}
\begin{center}
\renewcommand{\arraystretch}{0.7}
\begin{tabular}{c|ccccc}
& \multicolumn{5}{c}{ Symbol} \\
State & 0 & 1 & $X$ & $Y$ & $B$ \\ \hline
$q_0$ & $(q_1,X,R)$ & ---         & ---         & $(q_3,Y,R)$ & ---         \\
$q_1$ & $(q_1,0,R)$ & $(q_2,Y,L)$ & ---         & $(q_1,Y,R)$ & ---         \\
$q_2$ & $(q_2,0,L)$ & ---         & $(q_0,X,R)$ & $(q_2,Y,L)$ & ---         \\
$q_3$ & ---         & ---         & ---         & $(q_3,Y,R)$ & $(q_4,B,R)$ \\
$q_4$ & ---         & ---         & ---         & ---         & ---         \\
\end{tabular}
\end{center}
\caption{\label{fig:example-delta}The function $\delta$.}
\end{figure}

\begin{figure}
\begin{center}
\scalebox{0.45}{\includegraphics{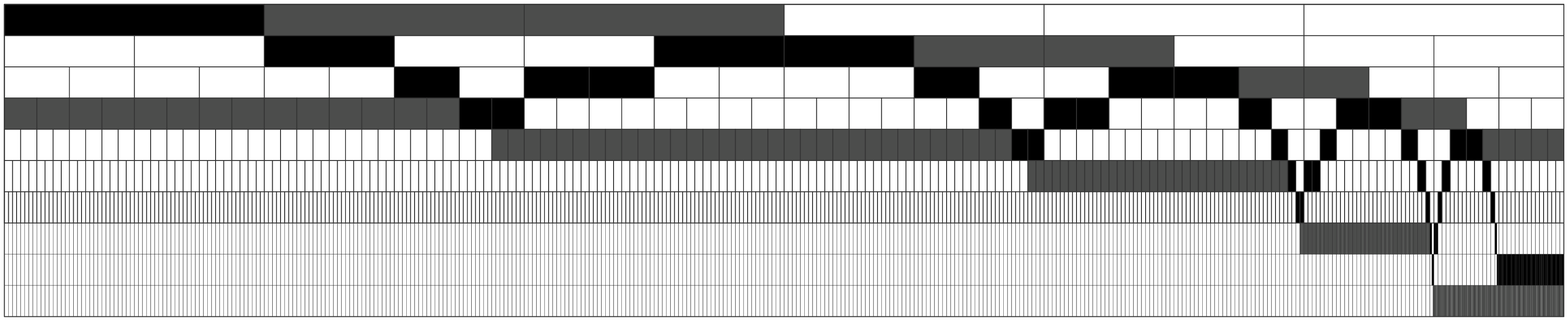}}
\caption{\label{fig:evolution} Space-time diagram of the computation of $A_M$ on input $01$.}
\end{center}
\end{figure}

We illustrate the working of $A_M$ by a simple example.
Let $L$ be the formal language consisting of strings with $n$ 0's, followed by $n$ 1's:
$L = \{0^n1^n | n \geq 1\}$.
A Turing machine that accepts this language is given by
$M = (\{q_0, q_1, q_2, q_3, q_4\}, \{0,1\}, \{0,1,X,Y,B\}, \delta, q_0, B, \{q_4\})$ \cite{hopcroft}
with the transition function depicted in Fig.~\ref{fig:example-delta}.
Note that $L$ is a context-free language, but $M$ will serve for demonstration purposes.
The computation of $M$ on input $01$ is given below:
\[
q_001 \vdash Xq_11  \vdash  q_2XY  \vdash  Xq_0Y  \vdash  XYq_3  \vdash XYBq_4.
\]
Fig.~\ref{fig:example-hyper-sca-2} depicts the computation of $A_M$ on the Turing machine input 01.
The first column of the table specifies the time in binary base.
$A_M$ performs 4 complete pulse zigzags and enters a final configuration in the fifth one after the Turing machine simulation has reached
the final state $q_4$.
Fig.~\ref{fig:evolution} depicts the space-time diagram of the computation.
It shows the position of the left and right delimiter (gray) and the position of the pulse (black).

\begin{figure}
\begin{center}
\scriptsize     {
\renewcommand{\arraystretch}{0.9}
\begin{tabular}{r|cccccccccccccc}
   &0 &1 & 2 & 3 & 4 & 5 & 6 & 7 & 8 & 9 \\ \hline
$0.00000000_2$ & $\overrightarrow{\lhd}$ & $\langle q_0,0 \rangle$ & $1$ & $\rhd$ & $\Box$ & $\Box$ & $\Box$ & $\Box$ & $\Box$ & $\Box$ \\
$1.00000000_2$ & $\lhd$ & $\overrightarrow{\langle q_0,0 \rangle}$ & $1$ & $\rhd$ & $\Box$ & $\Box$ & $\Box$ & $\Box$ & $\Box$ & $\Box$ \\
$1.10000000_2$ & $\lhd$ & $X$ & $\overrightarrow{\langle q_1,1 \rangle}$ & $\rhd$ & $\Box$ & $\Box$ & $\Box$ & $\Box$ & $\Box$ & $\Box$ \\
$1.11000000_2$ & $\lhd$ & $X$ & $\langle q_1,1 \rangle$ & $\rhd_\blacktriangleleft$ & $\Box$ & $\Box$ & $\Box$ & $\Box$ & $\Box$ & $\Box$ \\
$1.11100000_2$ & $\lhd$ & $X$ & $\langle q_1,1 \rangle$ & $\blacktriangleleft$ & $\rhd$ & $\Box$ & $\Box$ & $\Box$ & $\Box$ & $\Box$ \\
$10.00000000_2$ & $\lhd$ & $X$ & $\blacktriangleleft$ & $\langle q_1,1 \rangle$ & $\rhd$ & $\Box$ & $\Box$ & $\Box$ & $\Box$ & $\Box$ \\
$10.10000000_2$ & $\lhd$ & $\blacktriangleleft$ & $X$ & $\langle q_1,1 \rangle$ & $\rhd$ & $\Box$ & $\Box$ & $\Box$ & $\Box$ & $\Box$ \\
$11.00000000_2$ & $\Box$ & $\overrightarrow{\lhd}$ & $X$ & $\langle q_1,1 \rangle$ & $\rhd$ & $\Box$ & $\Box$ & $\Box$ & $\Box$ & $\Box$ \\
$11.10000000_2$ & $\Box$ & $\lhd$ & $\overrightarrow{X}$ & $\langle q_1,1 \rangle$ & $\rhd$ & $\Box$ & $\Box$ & $\Box$ & $\Box$ & $\Box$ \\
$11.11000000_2$ & $\Box$ & $\lhd$ & $\langle q_2,X \rangle$ & $\overrightarrow{Y}$ & $\rhd$ & $\Box$ & $\Box$ & $\Box$ & $\Box$ & $\Box$ \\
$11.11100000_2$ & $\Box$ & $\lhd$ & $\langle q_2,X \rangle$ & $Y$ & $\rhd_\blacktriangleleft$ & $\Box$ & $\Box$ & $\Box$ & $\Box$ & $\Box$ \\
$11.11110000_2$ & $\Box$ & $\lhd$ & $\langle q_2,X \rangle$ & $Y$ & $\blacktriangleleft$ & $\rhd$ & $\Box$ & $\Box$ & $\Box$ & $\Box$ \\
$100.00000000_2$ & $\Box$ & $\lhd$ & $\langle q_2,X \rangle$ & $\blacktriangleleft$ & $Y$ & $\rhd$ & $\Box$ & $\Box$ & $\Box$ & $\Box$ \\
$100.01000000_2$ & $\Box$ & $\lhd$ & $\blacktriangleleft$ & $\langle q_2,X \rangle$ & $Y$ & $\rhd$ & $\Box$ & $\Box$ & $\Box$ & $\Box$ \\
$100.10000000_2$ & $\Box$ & $\Box$ & $\overrightarrow{\lhd}$ & $\langle q_2,X \rangle$ & $Y$ & $\rhd$ & $\Box$ & $\Box$ & $\Box$ & $\Box$ \\
$100.11000000_2$ & $\Box$ & $\Box$ & $\lhd$ & $\overrightarrow{\langle q_2,X \rangle}$ & $Y$ & $\rhd$ & $\Box$ & $\Box$ & $\Box$ & $\Box$ \\
$100.11100000_2$ & $\Box$ & $\Box$ & $\lhd$ & $X$ & $\overrightarrow{\langle q_0,Y \rangle}$ & $\rhd$ & $\Box$ & $\Box$ & $\Box$ & $\Box$ \\
$100.11110000_2$ & $\Box$ & $\Box$ & $\lhd$ & $X$ & $\langle q_0,Y \rangle$ & $\rhd_B$ & $\Box$ & $\Box$ & $\Box$ & $\Box$ \\
$100.11111000_2$ & $\Box$ & $\Box$ & $\lhd$ & $X$ & $\langle q_0,Y \rangle$ & $B$ & $\rhd_\blacktriangleleft$ & $\Box$ & $\Box$ & $\Box$ \\
$100.11111100_2$ & $\Box$ & $\Box$ & $\lhd$ & $X$ & $\langle q_0,Y \rangle$ & $B$ & $\blacktriangleleft$ & $\rhd$ & $\Box$ & $\Box$ \\
$101.00000000_2$ & $\Box$ & $\Box$ & $\lhd$ & $X$ & $\langle q_0,Y \rangle$ & $\blacktriangleleft$ & $B$ & $\rhd$ & $\Box$ & $\Box$ \\
$101.00010000_2$ & $\Box$ & $\Box$ & $\lhd$ & $X$ & $\blacktriangleleft$ & $\langle q_0,Y \rangle$ & $B$ & $\rhd$ & $\Box$ & $\Box$ \\
$101.00100000_2$ & $\Box$ & $\Box$ & $\lhd$ & $\blacktriangleleft$ & $X$ & $\langle q_0,Y \rangle$ & $B$ & $\rhd$ & $\Box$ & $\Box$ \\
$101.01000000_2$ & $\Box$ & $\Box$ & $\Box$ & $\overrightarrow{\lhd}$ & $X$ & $\langle q_0,Y \rangle$ & $B$ & $\rhd$ & $\Box$ & $\Box$ \\
$101.01100000_2$ & $\Box$ & $\Box$ & $\Box$ & $\lhd$ & $\overrightarrow{X}$ & $\langle q_0,Y \rangle$ & $B$ & $\rhd$ & $\Box$ & $\Box$ \\
$101.01110000_2$ & $\Box$ & $\Box$ & $\Box$ & $\lhd$ & $X$ & $\overrightarrow{\langle q_0,Y \rangle}$ & $B$ & $\rhd$ & $\Box$ & $\Box$ \\
$101.01111000_2$ & $\Box$ & $\Box$ & $\Box$ & $\lhd$ & $X$ & $Y$ & $\overrightarrow{\langle q_3,B \rangle}$ & $\rhd$ & $\Box$ & $\Box$ \\
$101.01111100_2$ & $\Box$ & $\Box$ & $\Box$ & $\lhd$ & $X$ & $Y$ & $\langle q_3,B \rangle$ & $\rhd_B$ & $\Box$ & $\Box$ \\
$101.01111110_2$ & $\Box$ & $\Box$ & $\Box$ & $\lhd$ & $X$ & $Y$ & $\langle q_3,B \rangle$ & $B$ & $\rhd_\blacktriangleleft$ & $\Box$ \\
$101.01111111_2$ & $\Box$ & $\Box$ & $\Box$ & $\lhd$ & $X$ & $Y$ & $\langle q_3,B \rangle$ & $B$ & $\blacktriangleleft$ & $\rhd$ \\
$101.10000000_2$ & $\Box$ & $\Box$ & $\Box$ & $\lhd$ & $X$ & $Y$ & $\langle q_3,B \rangle$ & $\blacktriangleleft$ & $B$ & $\rhd$ \\
$101.10000100_2$ & $\Box$ & $\Box$ & $\Box$ & $\lhd$ & $X$ & $Y$ & $\blacktriangleleft$ & $\langle q_3,B \rangle$ & $B$ & $\rhd$ \\
$101.10001000_2$ & $\Box$ & $\Box$ & $\Box$ & $\lhd$ & $X$ & $\blacktriangleleft$ & $Y$ & $\langle q_3,B \rangle$ & $B$ & $\rhd$ \\
$101.10010000_2$ & $\Box$ & $\Box$ & $\Box$ & $\lhd$ & $\blacktriangleleft$ & $X$ & $Y$ & $\langle q_3,B \rangle$ & $B$ & $\rhd$ \\
$101.10100000_2$ & $\Box$ & $\Box$ & $\Box$ & $\Box$ & $\overrightarrow{\lhd}$ & $X$ & $Y$ & $\langle q_3,B \rangle$ & $B$ & $\rhd$ \\
$101.10110000_2$ & $\Box$ & $\Box$ & $\Box$ & $\Box$ & $\lhd$ & $\overrightarrow{X}$ & $Y$ & $\langle q_3,B \rangle$ & $B$ & $\rhd$ \\
$101.10111000_2$ & $\Box$ & $\Box$ & $\Box$ & $\Box$ & $\lhd$ & $X$ & $\overrightarrow{Y}$ & $\langle q_3,B \rangle$ & $B$ & $\rhd$ \\
$101.10111100_2$ & $\Box$ & $\Box$ & $\Box$ & $\Box$ & $\lhd$ & $X$ & $Y$ & $\overrightarrow{\langle q_3,B \rangle}$ & $B$ & $\rhd$ \\
$101.10111110_2$ & $\Box$ & $\Box$ & $\Box$ & $\Box$ & $\lhd$ & $X$ & $Y$ & $B$ & $\overrightarrow{\langle q_4,B \rangle}$ & $\rhd$ \\
\end{tabular}
}
\end{center}
\caption{\label{fig:example-hyper-sca-2}A computation of $A_M$ on input $01$.}
\end{figure}

\subsection{Proof}

We split the proof that $A_M$ is a hypercomputer into several steps.
We first show that the block transformations are well-defined and the pulse is preserved during evolution.
Afterwards we will prove that $A_M$ simulates $M$ correctly and we will show that $A_M$ represents an accelerating Turing machine.

Let $D = \{\overrightarrow{\lhd}, \rhd_B, \rhd_\blacktriangleleft, \overrightarrow{a}, \overrightarrow{\langle q,a \rangle} \}$ be the
set of elements that represent the downgoing pulse,  $U = \{\blacktriangleleft\}$ be the singleton that contains the upgoing pulse,
$P = D \cup U$, and
$R = Z \backslash P$ the remaining elements.
The following lemma states that the block transformations are unambiguous for the set of configurations we
consider and that the pulse is preserved during evolution.

\begin{lemma}
If the finite configuration $C$ contains exactly one element of $P$ then
the application of the block transformations \ref{tr:start-state} -- \ref{tr:up-lhd} is unambiguous and at most
one block transformation is applicable.
If a configuration $C^\prime$ with $C \vdash_{A_M} C^\prime$ exists, then $C^\prime$ contains exactly one
element of $P$ as well.
\end{lemma}
\begin{proof}
Note that the domains of all block transformations are pairwise disjoint.
This ensures that for all pairs $z_1z_2$ in $Z \times Z$ at most one block transformation is applicable.
Block transformations \ref{tr:start-state} -- \ref{tr:new-blank} are all subsets or elements of
$(D \times R) \times (R \times D)$,
block transformation \ref{tr:reflection-right} is element of $(D \times R) \times (U \times R)$,
block transformations \ref{tr:up} and \ref{tr:up-state} are subsets of
$(R \times U) \times (U \times R)$, and finally
block transformation \ref{tr:up-lhd} is element of $(R \times U) \times (R \times D)$.
Since the domain is either a subset of $D \times R$ or $R \times U$ the block transformations
are unambiguous if $C$ contains at most one element of $P$.
A configuration $C^\prime$ with $C \vdash_{A_M} C^\prime$ must be the result
of the application of exactly one block transformation.
Since each block transformation preserves the pulse, $C^\prime$ contains one pulse if and only if $C$ contains one.
\end{proof}

We introduce a mapping $\gamma$ that aims to decode a self-similar cellular automaton configuration into a Turing machine configuration.
Let $C$ be a finite configuration.
Then $\gamma(C)$ is the string in $(\Gamma \cup Q)^{*}$ that is formed of $C$ as following:
\begin{enumerate}
\item All elements in $\{\Box, \blacktriangleleft, \lhd, \overrightarrow{\lhd}, \rhd, \rhd_B, \rhd_\blacktriangleleft\}$ are omitted.
\item All elements of the form $\overrightarrow{a}$ are replaced by $a$ and all elements of the form
$\langle q,a  \rangle$ or $\overrightarrow{\langle q,a  \rangle}$ are replaced by the two symbols $q$ and $a$.
\item All other elements of the form $a$ are added as they are.
\item Leading or trailing blanks of the resulting string are omitted.
\end{enumerate}
The following lemma states that $A_M$ correctly simulates $M$.
\begin{lemma}
Let $c_1$, $c_2$ be configurations of $M$.
If $c_1 \vdash_M^* c_2$, then there exist two finite configurations $C_1$, $C_2$ of $A_M$
such that $\gamma(C_1) = c_1$, $\gamma(C_2) = c_2$, and $C_1 \vdash_{A_M}^* C_2$.
Especially if the initial configuration $C_0$ of $A_M$ satisfies  $\gamma(C_0) = c_1$,
then there exists a finite configuration $C_2$ of $A_M$, such that $\gamma(C_2) = c_2$
 and $C_0 \vdash_{A_M}^* C_2$.
\end{lemma}
\begin{proof}
If $c_1$ has the form $a_1 \ldots a_n q$ we consider without loss of generality $a_1 \ldots a_n q B$.
Therefore let $c_1 = a_1 \ldots a_{i-1} q a_i \ldots a_n$.
If $i < n$ or $i = n$ and $\delta_D(q, a_n) = L$ we choose
 $C_1 = \overrightarrow{\lhd}  a_1 \ldots a_{i-1} \langle q,a_i \rangle a_{i+1} \ldots a_n \rhd$.
 If $i = n$ and $\delta_D(q, a_n) = R$ we insert an additional blank:
 $C_1 = \overrightarrow{\lhd}  a_1 \ldots a_{n-1} \langle q,a_n \rangle  B \rhd$.
 In any case $\gamma(C_1)=c_1$ holds.
We show the correctness of the simulation by calculating a complete zigzag of the pulse for the
start configuration:
$\overrightarrow{\lhd}  a_1 \ldots a_{i-1} \langle q,a_i \rangle a_{i+1} \ldots a_n \rhd$.
The number of the block transformation that is applied, is written above the derivation symbol.
We split the zigzag up into three phases.

\begin{enumerate}
\item Pulse moves down from the left delimiter to the left neighbor cell of the simulated head.

For $i > 1$ we obtain
\begin{equation}
\begin{array}{l}
\overrightarrow{\lhd}  a_1 \ldots a_{i-1} \langle q,a_i \rangle a_{i+1} \ldots a_n \rhd
\stackrel{(\ref{tr:start})}{\vdash_{A_M}}
\lhd  \overrightarrow{a_1} \ldots a_{i-1} \langle q,a_i \rangle a_{i+1} \ldots a_n \rhd
\stackrel{(\ref{tr:down})}{\vdash_{A_M}} \\
\lhd  a_1 \overrightarrow{a_2} \ldots a_{i-1} \langle q,a_i \rangle a_{i+1} \ldots a_n \rhd
\stackrel{(\ref{tr:down})}{\vdash_{A_M}} \ldots \stackrel{(\ref{tr:down})}{\vdash_{A_M}}
\lhd  a_1 \ldots \overrightarrow{a_{i-1}} \langle q,a_i \rangle a_{i+1} \ldots a_n \rhd.
\label{der:start}
\end{array}
\end{equation}

If $i = 1$ the pulse piggybacked by the left delimiter $\overrightarrow{\lhd}$ is already
in the left neighbor cell of the head and this phase is omitted.

\item Downgoing pulse passes the head.

If in the beginning of the zigzag the head was to the right of the left delimiter then
\begin{equation}
\overrightarrow{\lhd} \langle q,a_1 \rangle a_{2} \ldots a_n \rhd
\stackrel{(\ref{tr:start-state})}{\vdash_{A_M}}
\lhd \overrightarrow{\langle q,a_1 \rangle} a_{2} \ldots a_n \rhd.
\end{equation}
If $\delta_D(q,a_1)=L$ no further block transformation is applicable and the configuration is final.
The case $\delta_D(q,a_1)=R$ will be handled later on.
We now continue the derivation~\ref{der:start}.
If $\delta(q,a_i) = (p, b, L)$ then
\begin{equation}
\lhd  a_1 \ldots \overrightarrow{a_{i-1}} \langle q,a_i \rangle a_{i+1} \ldots a_n \rhd
\stackrel{(\ref{tr:left-1})}{\vdash_{A_M}}
\lhd  a_1 \ldots \langle p,a_{i-1} \rangle \overrightarrow{b}  a_{i+1} \ldots a_n \rhd.
\label{der:tm-step-left}
\end{equation}

If $\delta(q,a_i) = (p, b, R)$ then
\begin{equation}
\lhd  a_1 \ldots \overrightarrow{a_{i-1}} \langle q,a_i \rangle a_{i+1} \ldots a_n \rhd
\stackrel{(\ref{tr:down-to-head})}{\vdash_{A_M}}
\lhd  a_1 \ldots a_{i-1} \overrightarrow{\langle q,a_i \rangle} a_{i+1} \ldots a_n \rhd.
\end{equation}
We distinguish two cases: $i < n$ and $i = n$.
If $i < n$ then
\begin{equation}
\lhd  a_1 \ldots a_{i-1} \overrightarrow{\langle q,a_i \rangle} a_{i+1} \ldots a_n \rhd
\stackrel{(\ref{tr:right-2})}{\vdash_{A_M}}
\lhd  a_1 \ldots a_{i-1} b \overrightarrow{\langle p,a_{i+1} \rangle} a_{i+2} \ldots a_n \rhd.
\label{pulse-passed}
\end{equation}
If the next steps of $M$ are moving the head again to the right, block transformation
\ref{tr:right-2} will repeatedly applied, till the head changes its direction or till the
head is left of the right delimiter $\rhd$.
If the Turing machine $M$ changes its direction before the right delimiter is reached, we obtain
\begin{equation}
\lhd  a_1 \ldots a_{i-1} b_1 \ldots b_j \overrightarrow{\langle r ,a_{k} \rangle} a_{k+1} \ldots a_n \rhd
\stackrel{(\ref{tr:left-no-move})}{\vdash_{A_M}}
\lhd  a_1 \ldots a_{i-1} b_1 \ldots b_j \langle r ,a_{k} \rangle \overrightarrow{a_{k+1}} \ldots a_n \rhd
\end{equation}
or if the direction change happens just before the right delimiter then
\begin{equation}
\lhd  a_1 \ldots a_{i-1} b_1 \ldots b_j \overrightarrow{\langle r ,a_{n} \rangle} \rhd
\stackrel{(\ref{tr:down-state-right-delimiter})}{\vdash_{A_M}}
\lhd  a_1 \ldots a_{i-1} b_1 \ldots b_j \langle r ,a_{n} \rangle \rhd_\blacktriangleleft.
\label{der:right}
\end{equation}
If $i=n$ or if the right-moving head hits the right delimiter the derivation has the following form
\begin{equation}
\lhd  a_1 \ldots a_{n-1} \overrightarrow{\langle q,a_n \rangle}  \rhd
\stackrel{(\ref{tr:down-state-right-delimiter-blank})}{\vdash_{A_M}}
\lhd  a_1 \ldots a_{n-1} \langle q,a_n \rangle  \rhd_B
\stackrel{(\ref{tr:new-blank})}{\vdash_{A_M}}
\lhd  a_1 \ldots a_{n-1} \langle q,a_n \rangle  B \rhd_\blacktriangleleft,
\label{der:new-blank}
\end{equation}
which inserts a blank to the right of the simulated head.

\item Downgoing pulse is reflected and moves up.

We proceed from configurations of the form
$\lhd  c_1 \ldots c_{i-1} \langle p,c_{i} \rangle \overrightarrow{c_{i+1}} \ldots c_n \rhd$. Then
\begin{equation}
\begin{array}{l}
\lhd  c_1 \ldots c_{i-1} \langle p,c_{i} \rangle \overrightarrow{c_{i+1}} \ldots c_n \rhd
\stackrel{(\ref{tr:down})}{\vdash_{A_M}} \ldots \stackrel{(\ref{tr:down})}{\vdash_{A_M}}
\lhd  c_1 \ldots c_{i-1} \langle p,c_{i} \rangle c_{i+1} \ldots \overrightarrow{c_n} \rhd
\stackrel{(\ref{tr:down-a-rhd})}{\vdash_{A_M}} \\
\lhd  c_1 \ldots c_{i-1} \langle p,c_{i} \rangle c_{i+1} \ldots  c_n \rhd_\blacktriangleleft
\stackrel{(\ref{tr:reflection-right})}{\vdash_{A_M}}
\lhd  c_1 \ldots c_{i-1} \langle p,c_{i} \rangle c_{i+1} \ldots  c_n \blacktriangleleft \rhd
\stackrel{(\ref{tr:up})}{\vdash_{A_M}} \ldots \stackrel{(\ref{tr:up})}{\vdash_{A_M}} \\
\lhd  c_1 \ldots c_{i-1} \langle p,c_{i} \rangle \blacktriangleleft c_{i+1} \ldots  c_n  \rhd
\stackrel{(\ref{tr:up-state})}{\vdash_{A_M}}
\lhd  c_1 \ldots c_{i-1} \blacktriangleleft \langle p,c_{i} \rangle  c_{i+1} \ldots  c_n  \rhd
\stackrel{(\ref{tr:up})}{\vdash_{A_M}} \ldots \stackrel{(\ref{tr:up})}{\vdash_{A_M}} \\
\lhd  \blacktriangleleft c_1 \ldots c_{i-1} \langle p,c_{i} \rangle  c_{i+1} \ldots  c_n  \rhd
\stackrel{(\ref{tr:up-lhd})}{\vdash_{A_M}}
\overrightarrow{\lhd}  c_1 \ldots c_{i-1} \langle p,c_{i} \rangle  c_{i+1} \ldots  c_n  \rhd,
\end{array}
\label{der:up}
\end{equation}
which finishes the zigzag.
Note that the continuation of derivations \ref{der:right} and \ref{der:new-blank} is handled by the
later part of derivation \ref{der:up}.
We also remark that the zigzag has shifted the whole configuration one cell downwards.

\end{enumerate}

All block transformations except transformations \ref{tr:right-2} and
\ref{tr:left-1} keep the $\gamma$-value of the configuration unchanged.
Block transformations \ref{tr:right-2} and \ref{tr:left-1} correctly simulate
one step in the calculation of the Turing machine $M$:
if
$C \stackrel{(\ref{tr:right-2}) or (\ref{tr:left-1})}{\vdash_{A_M}} C^\prime$, $\gamma(C)=c$, and $\gamma(C^\prime)=c^\prime$
then $c \vdash_M c^\prime$.
Let $C_1^\prime$ be the resulting configuration of the zigzag.
We conclude that $\gamma(C_1) \vdash_M^* \gamma(C_1^\prime)$ holds.
We have chosen $C_1$ in such a way that at least one step of $M$ is performed, if $M$ does not halt, either by
block transformation \ref{tr:right-2} or \ref{tr:left-1}.
If $M$ does not halt the configuration after the zigzag is again of the form
$\overrightarrow{\lhd}  a_1 \ldots a_{i-1} \langle q,a_i \rangle a_{i+1} \ldots a_n \rhd$.
The case $i = n$ and $\delta_D(q,a_n) = R$ is excluded by derivation \ref{der:new-blank},
which inserts a blank to the right of the head, if $\delta_D(q,a_n) = R$.
This means that $C_1^\prime$ has the same form as $C_1$ and that any subsequent zigzag will  perform at
least one step of $M$ as well if $M$ does not halt.

In summary, we conclude that $A_M$ reaches after a finite number of zigzags a configuration $C_2$ such that
$\gamma(C_2) = c_2$.
On the other hand, if $M$ halts, $A_M$ enters a final configuration since derivations
\ref{der:tm-step-left} or \ref{pulse-passed} are not applicable anymore
and the pulse cannot cross the simulated head.
Since we have chosen $C_0$ to be of the same form
as $C_1$ in the beginning of the proof, the addendum of the lemma regarding the initial configuration is true.
\end{proof}

Next, the time behavior of the self-similar cellular automaton $A_M$ will be investigated.
\begin{lemma}
Let $C=\overrightarrow{\lhd}  a_1 \ldots a_{i-1} \langle q,a_i \rangle a_{i+1} \ldots a_n \rhd$
be a finite configuration of $A_M$ that starts in cell $k$.
If $M$ does not halt, the zigzag of the pulse takes 3 cycles of cell $k$ and $A_M$
is afterwards in a finite configuration
$C^\prime=\overrightarrow{\lhd}  b_1 \ldots b_{j-1} \langle p,b_j \rangle b_{j+1} \ldots b_m \rhd$
that starts in cell $k + 1$.
\end{lemma}
\begin{proof}
Without loss of generality, we assume that the finite configuration starts in cell 0.
We follow the zigzag of the pulse, thereby tracking all times,
compare with Fig.~\ref{fig:example-hyper-sca-2} and Fig.~\ref{fig:evolution}.
The pulse reaches at time 1 cell 1,
and at time $\sum_{i=0}^1 2^{-i}$ cell 2.
In general, the downgoing pulse reaches cell $r$ in time $\sum_{i=0}^{r-1} 2^{-i}$.
At time $\sum_{i=0}^{n+1} 2^{-i}$ the cell $n+2$ changes to $\rhd_\blacktriangleleft$
which marks the reversal of direction of the pulse.
The next configuration change ($\rhd_\blacktriangleleft \Box \mapsto \blacktriangleleft \rhd$)
occurs at $\sum_{i=0}^{n+1} 2^{-i} + 2^{-(n+1)} = 2$.
The pulse $\blacktriangleleft$ reaches cell $n+1$ in time $2 + 2^{-(n+1)}$ and in general
cell $r$ in time $2 + 2^{-r}$.
The final configuration change of the zigzag ($\lhd  \blacktriangleleft \mapsto \Box  \overrightarrow{\lhd}$)
that marks also the beginning of a new pulse zigzag occurs synchronously in cell 0 and cell 1 at time 3.
We remark that the overall time of the pulse zigzag remains unchanged if the
simulated head inserts a blank between the two delimiters.
\end{proof}

\begin{theorem}
\label{th-rca}
If $M$ halts on $w$ and $A_M$ is initialized with $C_0(w)$ then $A_M$ enters a final
configuration in a time less than 6 cycles of cell 0, containing the result of the calculation between the left
and right delimiter.
If $M$ does not halt, $A_M$ enters after 6 cycles of cell 0 the final configuration that consists
of an infinite string of the quiescent element: $\Box^\infty$.
\end{theorem}
\begin{proof}
$A_M$ needs 3 cycles of cell 0 to perform the first zigzag of the pulse.
After the 3 cycles the configuration is shifted one cell downwards, starting now in cell 1.
The next zigzag takes 3 cycles of cell 1 which are 3/2 cyles of cell 0, and so on.
Each zigzags performs at least one step of the Turing machine $M$, if $M$ does not halt.
We conclude that if $M$ halts, $A$ enters a final configuration in a time less than
$\sum_{i=0}^\infty 3/2^{i} = 6$ cycles of cell 0.
If $M$ does not halt, the zigzag disappears in infinity after 6 cycles of cell 0 leaving a trail of $\Box$'s behind.
\end{proof}

If $M$ is a universal Turing machine, we immediately obtain the following result, which proves that $A_M$ is a hypercomputer
for certain Turing machines $M$.
\begin{cor}
Let $M_U$ be a universal Turing machine. Then $A_{M_U}$ solves the halting problem for Turing machines.
\end{cor}
\begin{proof}
Initialize $A_{M_U}$ with an encoded Turing machine $M$ and an input word $w$.
Then $A_M$ enters a final configuration with the result of $M$ on $w$ in less than 6 cycles of cell 0 if and only if $M$ halts.
\end{proof}

In the current form of Turing machine simulation the operator has to scan a potentially unlimited number of cells to determine whether
$M$ has halted or not, which limits its practical value.
If $M$ has halted, we would like to propagate at least this fact back to the upper cells.
The following obvious strategy fails in a subtle way.
Add a rule to $A_M$ that whenever $\langle q, a \rangle$ has no next move, replaces it by the new symbol $H$.
Add the rule $f(?, ?, H, ?) = H$ to $A_M$
that propagates $H$ upwards to cell $0$.
The propagation upwards is only possible if we change also the block transformation \ref{tr:up-lhd}  to
$\lhd  \blacktriangleleft \mapsto \Diamond  \overrightarrow{\lhd}$,
thereby introducing a new symbol $\Diamond$ that is not subject
of the short-circuit evaluation.
The last point, even if necessary, causes the strategy to fail, since if $A_M$ does not halt,
$A_M$ is after 6 cycles in the configuration $\Diamond^\infty$ that
leads to indeterministic behavior of $A_M$.
This is in so far problematic, since we can not be sure whether a state $H$ in cell $0$
is really the outcome of a halting Turing machine or the result of indeterministic behavior.
Instead of enhancing the self-similar cellular automaton model, we will introduce in the next section a computing model
that is computational equivalent for finite computations, but avoids indeterminism for infinite
computations.

\section{Self-similar Petri nets}
\label{chap:petri}

The evolution of a cellular automaton as well as the evolution of a self-similar cellular automaton depends on an extrinsic clock representing a global time that
triggers the state changes.
Since a self-similar cellular automaton cannot halt, a self-similar cellular automaton is forced to perform a state change, even if no state
with a causal relationship to the previous one exists, leading to indeterministic behavior,
as described in the introduction.
In this section, we present a model based on Petri nets, the self-similar Petri nets, with a close resemblance to self-similar cellular automata.
Even though Petri nets in general are not deterministic, there exist subclasses that are.
As  will be shown below, self-similar Petri Nets are deterministic.
They are also capable of hypercomputing, but compared to self-similar cellular automata, their behavior differ in the limit.
Whereas a self-similar cellular automaton features indeterministic behavior, the self-similar Petri net halts.

\subsection{Petri nets}

C.A.~Petri introduced Petri nets in the 1960s to study asynchronous computing systems.
They are now widely used to describe and study information processing systems
that are characterized as being concurrent, asynchronous, distributed, parallel, nondeterministic,
and/or stochastic.
It is interesting to note that very early, and clearly ahead of its time, Petri investigated
the connections between physical and computational processes, see e.g., Ref.~\cite{petri-82}.
In what follows, we give a brief introduction to Petri nets to define the terminology.
For a more comprehensive treatment we refer to the literature; e.g., to Ref.~\cite{Murata89}.

\begin{defn}[Petri Net]
A Petri net is a directed, weighted, bipartite graph consisting of two kinds of nodes,
called places and transitions.
The weight $w(p,t)$ is the weight of the arc from place $p$ to transition $t$,
$w(t,p)$ is the weight of the arc from transition $t$ to place $p$.
A marking assigns to place $p$ a nonnegative integer $k$, we say that $p$ is marked with
$k$ tokens.
If a place $p$ is connected with a transition $t$ by an arc that goes from $p$ to $t$,
$p$ is an input place of $t$, if the arc goes from $t$ to $p$, $p$ is an output place.
A Petri net is changed according to the following transition (firing) rule:
\begin{enumerate}
\item
a transition $t$ may fire if each input place $p$ of $t$ is marked with at least
$w(p,t)$ tokens, and
\item
a firing of an enabled transition $t$ removes $w(p,t)$ tokens from each input place $p$ of $t$,
and adds $w(t,p)$ tokens to each output place $p$ of $t$.
\end{enumerate}
Formally, a Petri net $N$ is a tuple $N=(P,T,F,W,M_0)$ where
$P$ is the set of places, $T$ is the set of transitions,
$F \subseteq (P \times T) \cup (T \times P)$ is the set of arcs,
$W: F \rightarrow \mathbb{N}$ is the weight function,
and $M_0: P \rightarrow \mathbb{N}$ is the initial marking.
\end{defn}

In graphical representation, places are drawn as circles and transitions as boxes.
If a place is input place of more than one transition, the Petri net becomes in general
indeterministic, since a token in this place might enable more than one transition,
but only one can actually fire and consume the token.
The subclass of Petri nets given in the following definition avoids these conflicts and is therefore deterministic.
In a standard Petri net, tokens are indistinguishable.
If the Petri net model is extended so that the tokens can hold values, the Petri net is called a colored Petri net.

\begin{defn}[Marked Graph and Colored Petri Net]
A marked graph is a Petri Net such that each place has exactly one input transition and
exactly one output transition.
A colored Petri net is a Petri net where each token has a value.
\end{defn}

\subsection{Self-similarity}

It is well-known that cellular automata can be modeled as colored Petri Nets.
To do this, each cell of the cellular automaton is replaced by a transition and a place for each neighbor.
The neighbor transitions send their states as token values to their output places, which
are the input places of the transition under consideration.
The transition consumes the tokens, calculates the  new state, and send its state back to
its neighbors.
A similar construction can be done for self-similar cellular automata, leading
to the class of self-similar Petri nets.

\begin{figure}
\begin{center}
\includegraphics[scale=0.6]{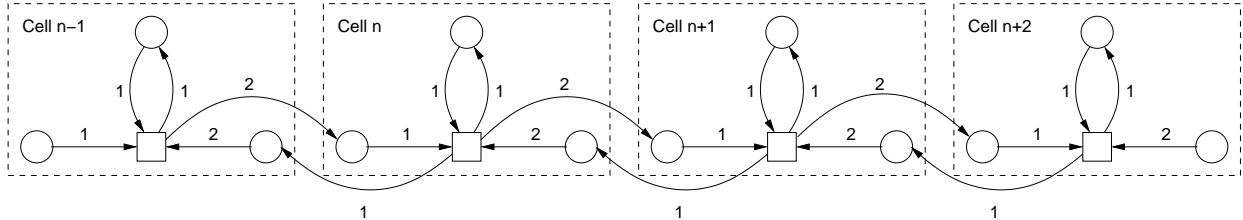}
\caption{\label{petri} Underlying graph of a self-similar Petri net.}
\end{center}
\end{figure}

\begin{defn}[Self-similar Petri Net]
A {\em self-similar Petri net} is a colored Petri net with some extensions.
A self-similar Petri net has the underlying graph partitioned into cells that is depicted in Fig.~\ref{petri}.
We denote the transition of cell $n$ by $t(n)$, the place to the left of the transition by
$p_l(n)$, the place to the right of the transition by $p_r(n)$ and the central place, in the figure the place above the transition, by $p_c(n)$.
Let $Z$ be a finite set, the state set, $q \in Z$ be the quiescent state, and $f$ be a (partial) function
 $Z^4 \times \{0,1\} \rightarrow Z$.
The set $V=Z \cup (\{0,1\} \times Z)$ is the value set of the tokens.
Tokens are added to a place and consumed from the place according to a first-in first-out order.
Initially, the self-similar Petri net starts with a finite number of cells $0, 1, \ldots, n$, and is allowed to grow to the right.
The notation $p \leftarrow z$ defines the following action:
create a token with value $z$ and add it to place $p$.
The firing rule for a transition in cell $n$ of a self-similar Petri net extends the firing rule of a standard Petri net in the following way:
\begin{enumerate}
\item If the transition $t(n)$ is enabled, the transition
removes token $\mathit{Tk}_l$ from place $p_l(n)$, token $\mathit{Tk}_c$ from $p_c(n)$ and
tokens $\mathit{Tk}_{r1}, \mathit{Tk}_{r2}$ from $p_r(n)$.
The value of token $\mathit{Tk}_{l}$ shall be of the form $(\mathit{coupled}, z_l)$ in $V=\{0,1\} \times Z$,
the other token values $z_c, z_{r1}$ and $z_{r2}$ shall be in $Z$.
If the tokens do not conform, the behavior of the transition is undefined.
\item
The transition calculates $z = f(z_l, z_c, z_{r1}, z_{r2}, \mathit{coupled})$.
\item
\emph{(Left boundary cell)}
If $n = 0$ then
$p_l(0) \leftarrow (\neg coupled, q)$, $p_c(0) \leftarrow z$, $p_l(1) \leftarrow (0, z)$, $p_l(1) \leftarrow (1,z)$.
\item
\emph{(Inner cell)}
If $n > 0$ and $n$ is not the highest index, then:
$p_r(n-1) \leftarrow z$, $p_c(n) \leftarrow z$, $p_l(n+1) \leftarrow (0, z)$, $p_l(n+1) \leftarrow (1, z)$.
\item \emph{(Right boundary cell)}
If $n$ is the highest index then:
\begin{enumerate}
\item \emph{(Quiescent state)}
\label{firing-rule-quiescent}
If $z = q$ then
$p_r(n-1) \leftarrow q$, $p_c(n) \leftarrow q$, $p_r(n) \leftarrow q$, $p_r(n) \leftarrow q$
\item \emph{(New cell allocation)}
If $z \neq q$ then a new cell $n + 1$ is created and connected to cell $n$.
Furthermore: $p_r(n-1) \leftarrow z$, $p_c(n) \leftarrow z$, $p_r(n) \leftarrow q$,
$p_l(n+1) \leftarrow (0, z)$, $p_l(n+1) \leftarrow (1, z)$,
$p_c(n + 1) \leftarrow q$, $p_r(n + 1) \leftarrow q$, $p_r(n + 1) \leftarrow q$.
\end{enumerate}
\end{enumerate}
Formally, we denote the self-similar Petri net by a tuple $N = (Z, f)$.
\end{defn}

A self-similar Petri net is a marked graph and therefore deterministic.
The initial markup is chosen in such a way that initially only the rightmost transition is enabled.

\begin{defn}(Initial markup)
Let $a_0 a_1 \ldots a_m$ be an input word in $Z^{m+1}$ and let
$N$ be a self-similar Petri net with $n$ cells, whereby $n > m + 1$.
The initial markup of the Petri net is as follows:
\begin{itemize}
\item $p_l(0) \leftarrow (0, q)$,
($p_l(i) \leftarrow (0, a_{i-1})$, $p_l(i) \leftarrow (1, a_{i-1})$) for $0 < i \leq m + 1$,
($p_l(i) \leftarrow (0, q)$, $p_l(i) \leftarrow (1, q)$) for  $i > m + 1$
\item $p_c(i) \leftarrow a_i$ for $i \leq m$, $p_c(i) \leftarrow q$ for $i > m$,
\item
$p_r(i) \leftarrow a_{i+1}$ for $i < m$, $p_r(i) \leftarrow q$ for $i \geq m$, and $p_r(n) \leftarrow q$.
\end{itemize}
\end{defn}

Note that the place $p_r(n)$ is initialized with two tokens.
We identify the state of a cell with the value of its $p_c$-token.
If $p_c$ is empty, because
the transition is in the process of firing, the state shall be the value of the last consumed token of $p_c$.

\begin{figure}
\begin{center}
\includegraphics[scale=0.8]{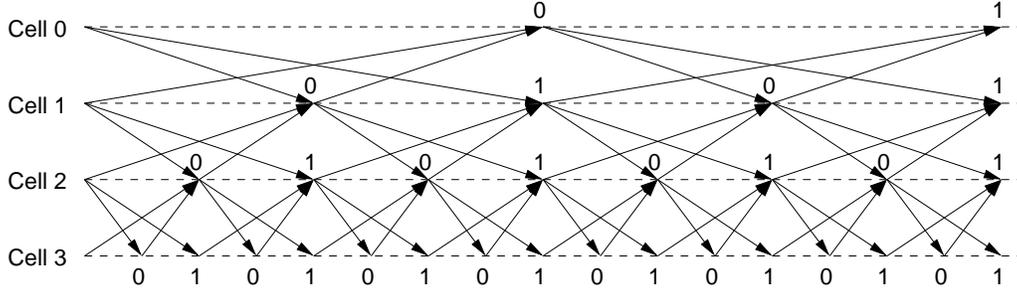}
\caption{\label{token-flow} Token flow in a self-similar Petri net.}
\end{center}
\end{figure}

Fig.~\ref{token-flow} depicts the token flow of a self-similar Petri net consisting of $4$ cells under the assumption
that the self-similar Petri net does not grow.
Tokens that are created and consumed by the same cell are not shown.
The numbers indicate whether the firing is uncoupled (0) or coupled (1).
The only transition that is enabled in the begin is $t(3)$, since $p_r(3)$ was initialized with 2 tokens.
The firing of $t(3)$ bootstraps the self-similar Petri net by adding a second token to $p_r(2)$, thereby enabling $t(2)$, and so on,
until all transitions have fired,
and the token flow enters periodic behavior.

\subsection{Comparison of self-similar cellular automata and self-similar Petri nets}

We now compare self-similar Petri nets with self-similar cellular automata.
We call a computation finite, if it involves either only a finite number of state updates of a self-similar cellular automaton, or
a finite number of transition firings of a self-similar Petri net, respectively.
\begin{lemma}
\label{lemma:comp-equivalence}
For finite computations, a dynamically growing self-similar cellular automaton $A = (Z, f)$ and a self-similar Petri net $N = (Z, f)$ are computationally
equivalent on a step-by-step basis if the start with the same number of cells and the same initial configuration.
\end{lemma}
\begin{proof}
Let $N$ be a self-similar Petri net which has $n$ cells initially.
For the sake of the proof consider an enhanced self-similar Petri net $N^\prime$ that is able to timestamp its token.
A token $\mathit{Tk}$ of $N^\prime$ does not hold only a value, but also a time interval.
We refer to the time interval of $\mathit{Tk}$ by $\mathit{Tk}.t$ and to the value of $\mathit{Tk}$ by $\mathit{Tk}.v$.
We remark that the timestamps serve only to compare the computations of a self-similar cellular automaton and a self-similar Petri net and do not imply any time behavior of the self-similar Petri net.
The firing rule of $N^\prime$ works as for $N$, but has an additional pre- and postprocessing step:
\begin{itemize}
\item \emph{(Preprocessing)}
Let $\mathit{Tk}_c$, $\mathit{Tk}_l$, $\mathit{Tk}_{r1}$, and $\mathit{Tk}_{r2}$ be the consumed token, where the alphabetical subscript denotes
the input place and the numerical subscript the order in which the tokens were consumed.
Calculate $t = (\mathit{Tk}_c.t)_\rightarrow$, where $\rightarrow$ is the inverse time operator of $\leftarrow$.
If $\mathit{Tk}_{r1}.t \neq t_\swarrow$ or $\mathit{Tk}_{r2}.t \neq t_\searrow$ or $\mathit{Tk}_{l}.t \neq t_\uparrow$ the firing fails
and the transition becomes permanently disabled.
\item \emph{(Postprocessing)}
For each created token $\mathit{Tk}$, set $\mathit{Tk}.t = t$.
\end{itemize}
The initial marking must set the $t$-field, otherwise the first transitions will fail.
For the initial tokens in cell $k$,
set $\mathit{Tk}_l.t = 2^{-k+1} \mathbbm{1}$ for both tokens in place $p_l$, $\mathit{Tk}_c.t = 2^{-k} \mathbbm{1}$,
and $\mathit{Tk}_d.t = 2^{-k-1} \mathbbm{1}$.
Set $\mathit{Tk}_d.t = 2^{-n-1} (1 + \mathbbm{1})$ for the second token in $p_r(n)$.
The firings of cell $k$ add tokens with timestamps $2^{-k} \mathbbm{1}, 2^{-k} (2 + \mathbbm{1}), 2^{-k} (3 + \mathbbm{1}) \ldots$
to the output place $p_c(k)$.
If transition $t(k)$ does not fail, the state function for the arguments $c=2^{-k} \mathbbm{1}$ and $t=2^{-k} (i + \mathbbm{1})$ is well-defined:
$s^\prime(c,t) = z$ if cell $k$ has produced or was initialized in place $p_r$ with a token $\mathit{Tk}$ with $\mathit{Tk}.t = t$ and $\mathit{Tk}.v = z$.
Let $s(c,t)$ be the state function of the scale-invariant cellular automaton $A$.
Due to the initialization, the two state functions are defined for the first $n$ cells and first time intervals $2^{-k} \mathbbm{1}$.
Assume that the values of $s$ and $s^\prime$ differ for some argument or that their domains are different.
Consider the first time interval $t_1$ where the difference occurs:
$s(c,t_1) \neq s^\prime(c,t_1)$, or exactly one of $s(c,t_1)$ or $s^\prime(c,t_1)$ is undefined.
If there is more than one time interval choose an arbitrary one of these.
Since $t_1$ was the first time interval where the state functions differ, we know that
$s(c_\uparrow, {t_1}_\uparrow) = s^\prime(c_\uparrow, {t_1}_\uparrow)$,
$s(c, {t_1}_\leftarrow) = s^\prime(c, {t_1}_\leftarrow)$,
$s(c_\swarrow, {t_1}_\swarrow) = s^\prime(c_\swarrow, {t_1}_\swarrow)$, and
$s(c_\swarrow, {t_1}_\searrow)= s^\prime(c_\swarrow, {t_1}_\searrow)$.
We handle the case that the values of the state functions are different or that $s^\prime$ is undefined for $(c,t_1)$ whereas $s$ is.
The other case ($s^\prime$ defined, but not $s$) can be handled analogously.
If $c = 2^{-k} \mathbbm{1}$, we conclude that tokens with timestamps ${t_1}_\uparrow$, ${t_1}_\leftarrow$, ${t_1}_\swarrow$, ${t_1}_\searrow$ were sent
to cell $k$, and no other tokens were sent afterwards to cell $k$, since the timestamps are created in
chronological order.
Hence, the precondition of the firing rule is satisfied and we conclude that $s(c,t_1) = s^\prime(c,t_1)$, which contradicts our assumption.
The allocation of new cells introduces some technicalities, but the overall strategy of going back in time
and concluding that the conditions for a state change or cell allocation were the same in both models works here also.
We complete the proof, by the simple observation that $N$ and $N^\prime$ perform the same computation.
\end{proof}
The proof can be simplified using the following more abstract argumentation.
A comparison of Fig.~\ref{token-flow} with Fig.~\ref{fig:timeops} shows that each computation step has in both models
the same causal dependencies.
Since both computers use the same rule to calculate the value of a cell, respectively the value of a token,
we conclude that the causal nets \cite{Levin81} of both computations are the same for a finite computation,
and therefore both computers yield the same output, in case the computation is finite.

\subsection{Timed self-similar Petri nets that hypercompute}

A large number of different approaches to introducing time concepts to
Petri nets have been proposed since the first extensions in the mid 1970s.
We do not delve into the depths of the different models, but instead,
define a very simple time schedule for the class of self-similar Petri nets.

\begin{defn}[Timed Self-similar Petri Net]
A timed self-similar Petri net is a self-similar Petri net that fires as soon as the transition is enabled and where a firing of an enabled transition $t(k)$ takes the time $2^{-k}$.
In the beginning of the firing, the tokens are removed from the input places, and at the end of the firing
the produced tokes of the firing are simultaneously entered into the output places.
\end{defn}

This time model can be satisfied if the cells of the timed self-similar Petri net  are arranged as the cells of a self-similar cellular automaton.
Under the assumption of a constant token speed, a firing time that is  proportional to the cell length,
and an appropriate unit of time
we yield  again cycle times of $2^{-k}$.

We now come back to the simulation of Turing machines and construct a hypercomputing timed self-similar Petri net, analogous to the
hypercomputing self-similar cellular automaton in section \ref{chap:hypercomputer}.
Let $M = (Q, \Sigma, \Gamma, \delta, q_0, B, F)$ be an arbitrary Turing machine.
Let $Z$ be the state set that we used in the simulation of a Turing machine by a self-similar cellular automaton, and let
$f$ the local rule that is defined by the block transformations \ref{tr:start-state} - \ref{tr:up-lhd},
without the short-circuit evaluation.
By Lemma \ref{lemma:comp-equivalence} we know that the timed self-similar Petri net $N_M = (Z, f)$ simulates $M$ correctly for a finite number of Turing machine steps.
Hence, if $M$ halts on input $w$, $N_M$ enters a final configuration in less than 6 cyles of cell 0.
We examine now the case that $M$ does not halt.
A pivotal difference between a self-similar cellular automaton and a self-similar Petri net is the ability of the latter one to halt on a computation.
This happens if all transitions of the self-similar Petri net are disabled.

\begin{lemma}
\label{lemma:apn-halting}
Let $M = (Q, \Sigma, \Gamma, \delta, q_0, B, F)$ be an arbitrary Turing machine and $w$ an input word in $\Sigma^*$.
If $M$ does not halt on $w$, the timed self-similar Petri net $N_M$ halts on $C_0(w)$ after 6 cycles of cell 0.
\end{lemma}
\begin{proof}
As long as the number of cells is finite, the boundary condition~\ref{firing-rule-quiescent} of the firing rule adds by each firing
two tokens to the $p_r$-place of the rightmost cell that successively enable all other transitions as well.
This holds no longer for the infinite case.
Let $M$ be a Turing machine, and $w$ an input word, such that $M$ does not halt on $w$.
We consider again the travel of the pulse zigzags down to infinity for the timed self-similar Petri net $N_M$ with
initial configuration $C_0(w)$, thereby tracking the marking of the $p_r$-places
for times after the zigzag has passed by.
The first states of cell $0$ are $\overrightarrow{\lhd}$, $\lhd$, $\lhd$, and $\Box$, including the initial one.
The state $\Box$ is the result of the firing at time 3, exhausting thereby the tokens in place $p_r(0)$.
At time 3 the left delimiter ($\overrightarrow{\lhd}$) of the pulse zigzag is now in cell 1.
Cell 1 runs from time 3 on through the same state sequence $\overrightarrow{\lhd}$, $\lhd$, $\lhd$, and $\Box$,
thereby adding in summary 4 tokens to $p_r(0)$.
After creating the token with value $\Box$, $p_r(1)$ is empty as well.
We conclude that after the zigzag has passed by a cell, the lower cell sends in summary 4 tokens to the upper cell,
till the zigzag has left the lower cell as well.
For each cell $k$ these four tokens in $p_r(k)$ enable two firings of cell $k$ thereby adding two tokens
to $p_r(k-1)$.
These two tokens of $p_r(k-1)$ enable again one firing of cell $k-1$ thereby adding one token to $p_r(k-2)$.
We conclude that each cell fires 3 times after the zigzag has passed by and that the final marking of each $p_r$ is one.
Hence, no $p_r$ has the necessary two tokens that enable the transition, therefore all transitions are disabled
and $N_M$ halts at time 6.
\end{proof}

Since $N_M$ halts for nonhalting Turing machines, there are no longer any obstacles that prevent
the construction of the proposed propagation of the halting state back to upper cells.
We replace block transformation \ref{tr:start-state} with the following two and add one new.

If $\delta(q,a) = (p,c,R)$ set
\begin{equation}
\overrightarrow{\lhd} \: \langle q, a \rangle \mapsto \lhd \:
\overrightarrow{\langle q, a \rangle}.
\label{tr:start-state2}
\end{equation}

If $\delta(q,a) = (p,c,L)$ or $\delta(q,a)$ is not defined set
\begin{equation}
\overrightarrow{\lhd} \: \langle q, a \rangle \mapsto \lhd \: H.
\label{tr:H}
\end{equation}

If $\delta(q,a)$ is not defined set
\begin{equation}
\overrightarrow{b} \: \langle q, a \rangle \mapsto b \: H.
\label{tr:down-to-head2}
\end{equation}

The following definition propagates the state $H$ up to cell $0$:
\begin{equation}
f(?, ?, H, ?) = H.
\label{eq:H-up}
\end{equation}
We denote the resulting timed self-similar Petri net by $\overline{N}_M$.
The following theorem makes use of the apparently paradoxical fact, that $\overline{N}_M$ halts if and only if the
simulated Turing machine does not halt.
\begin{theorem}
Let $M_U$ be a universal Turing machine. Then $\overline{N}_{M_U}$ solves the halting problem for Turing machines.
\end{theorem}
\begin{proof}
Consider a Turing machine $M$ and an input word $w$. Initialize $\overline{N}_{M_U}$ with $C_0(\langle M, w \rangle )$ where
$\langle M, w \rangle$ is the encoding of $M$ and $w$.
If $M$ does not halt on $w$, $\overline{N}_{M_U}$ halts at time 6 by Lemma~\ref{lemma:apn-halting}.
If $M$ halts on $w$, then one cell of $\overline{N}_{M_U}$ enters the state $H$ by
block transformation~\ref{tr:H} or~\ref{tr:down-to-head2} according to
Theorem~\ref{th-rca} and Lemma~\ref{lemma:comp-equivalence} and taking the changes in $f$ into account.
The mapping~\ref{eq:H-up} propagates $H$ up to cell 0.
An easy calculation shows that cell 0 is in state $H$, in time 7 or less.
\end{proof}

We have proven that $\overline{N}_{M_U}$ is indeed a hypercomputer without the deficiencies of the scale-invariant cellular automaton-based
hypercomputer. We end this section with two remarks.
The timed self-similar Petri net $N_M$ sends a flag back to the upper cells, if the simulated Turing machine halts.
Strictly speaking, this is not necessary, if the operator is able to recognize whether the timed self-similar Petri net has halted or not.
On the other hand, a similar construction is essential, if the operator is interested in the final tape content of the
simulated Turing machine.
Transferring the whole tape content of the simulated Turing machine upwards,
could be achieved by implementing a second pulse that performs an upwards-moving zigzag.
The construction is even simpler as the described one, since the tape content of the Turing machine becomes static as soon as the Turing machine halts.

The halting problem of Turing machines is not the only problem that can be solved by self-similar cellular automata, scale-invariant cellular automata, or timed self-similar Petri nets, but is unsolvable for Turing machines.
A discussion of other problems unsolvable by Turing machines and of techniques to solve them within infinite computing machines, can be found in
Davies~\cite{Davies01}.

\section{Summary}

We have presented two new computing models that implement the potential
infinite divisibility of physical configuration space.
These models are purely information theoretic and
do not take into account kinetic and other
effects.
With these provisos, it is possible, at least in
principle, to use the potential infinite
divisibility of space-time to perform hypercomputation,
thereby extending the algorithmic domain to hitherto unsolvable decision problems.

Both models are composed of elementary computation primitives.
The two models are closely related but are very different ontologically.
A cellular automaton depends on an {\em extrinsic} time requiring an {\em external} clock
and a rigid synchronization of its computing cells, whereas
a Petri net implements a causal relationship leading to an {\em intrinsic} concept of time.

Scale-invariant cellular automata as well as self-similar Petri nets are built in the same way from their primitive building blocks.
Each unit is recursively coupled with a sized-down copy of itself, potentially leading
to an infinite sequence of ever decreasing units.
Their close resemblance leads to a step-by-step equivalence of finite computations,
yet their ontological difference yields different behaviors for the for the case that the computation involves an infinite number of units:
a scale-invariant cellular automaton exhibits indeterministic behavior, whereas a self-similar Petri net halts.
Two supertasks which operate identically in the finite case but differ in
their limit is a puzzling observation which might question our present understanding of supertasks.
This may be considered an analogy to a theorem \cite{Specker49} in recursive analysis
about the existence of recursive monotone bounded sequences of rational numbers
whose limit is not a computable number.

One striking feature of both models is their scale-invariance.
The computational behavior of these models is therefore the first example for what might be called
scale-invariant or self-similar computing, which might be characterized by the property that
any computational space-time pattern can be arbitrary squeezed to finer and finer regions of space and time.

Although the basic definitions have been given, and elementary properties of these new models have been explored,
a great number of questions remain open for future research.
The construction of a hypercomputer was a first demonstration of the
extraordinary computational capabilities of these models.
Further investigations are necessary to determine their limits, and to relate them with the
emerging field of hypercomputation~\cite{2002-cal-pav,ord-2002,Davis-2004,Doria-2006,Davis-2006,potgieter-06,1011191}.
Another line of research would be the investigation of their phenomenological properties, analogous
to the statistical mechanics of cellular automata~\cite{wolfram83,wolfram-2002}.


\end{document}